\documentstyle[aps,preprint,prl]{revtex}
\tightenlines
\newcommand{\s}{\mbox{$\sigma$}}
\newcommand{\be}{\begin{equation}}
\newcommand{\ee}{\end{equation}}
\newcommand{\ua}{\mbox{$\uparrow$}}
\newcommand{\da}{\mbox{$\downarrow$}}
\newcommand{\k}{{\bf k}}

\begin{document}

\title{\Large{\bf Spin-triplet superconducting pairing 
due to local (Hund's rule, Dirac) exchange }}

\vskip0.5cm 

\author{ Jozef Spa\l{}ek$^*$  }

\address{Marian Smoluchowski Institute of Physics, Jagiellonian 
University, ulica Reymonta 4, \\
 30-059 Krak\'ow, Poland}

\maketitle

\vskip1.0cm

\begin{abstract}
We discuss general implications of the local spin-triplet pairing among 
fermions 
induced by  local ferromagnetic exchange, example of which is the Hund's 
rule coupling.  
The quasiparticle energy and their wave function are determined for the three
principal phases with the gap, which is  momentum independent. 
We utilize the Bogolyubov-Nambu-De Gennes approach, which in the case of 
triplet pairing in the two-band case 
leads to  the four-components wave function.   
Both gapless modes and those with an isotropic gap appear in the quasiparticle
spectrum.  
A striking analogy with the Dirac equation is briefly explored.  
This type of pairing is relevant to relativistic fermions as well, since it
reflects the fundamental discrete symmetry-particle interchange.   
A comparison with the local interband spin-singlet pairing  is also made.  

\end{abstract}

\vskip1.5cm

\noindent PACS Nos. 74.20Mn, 03.65.Pm


\noindent -----------------------------------------------------------

\noindent $^*)$ Electronic address: ufspalek@if.uj.edu.pl

\vskip0.5cm

\newpage

\noindent {\Large {\bf I. Introduction} }

\vskip0.5cm 

The discovery \cite{1} of superconductivity in the orbitally 
degenerate system 
$Sr_2RuO_4$, which is closely related to both ferromagnetic \cite{2}
$SrRuO_3$  
and antiferromagnetic and Mott insulating \cite{3} $Ca_2RuO_4$, poses a
question about the role of short-range Coulomb and exchange interactions in
stabilizing the spin-triplet superfluid state \cite{3}. 
In the case of Mott-Hubbard insulators the kinetic exchange 
interaction \cite{4} 
plays an essential role in stabilizing antiferromagnetism.  
This interaction is also instrumental in the form of {\it real space 
pairing} \cite{5} in driving the system close to the Mott-Hubbard 
boundary towards 
spin-singlet superconducting state.  
In the case of orbitally degenerate systems, the ferromagnetic \cite{6} 
and 
antiferromagnetic kinetic exchange interactions compete with each 
other \cite{7} for the number of electrons per atom $n >1$.  
Ferromagnetism (with a possible orbital ordering) usually wins \cite{8}  for
$n \rightarrow 1$, whereas the antiferromagnetism takes over when $n
\rightarrow d$, where $d$ is the orbital degeneracy.  
This type of competition should also be present in $Sr_2RuO_4$, in which $4d^4$
configuration of $Ru^{4+}$ contains two holes in $t_{2g}$ shell composed of
nominally triply degenerate $d_{\epsilon} = (d_{xy}, d_{yz}, d_{zx})$ 
orbitals.  
The two-dimensional antiferromagnetic spin fluctuations have been indeed
observed in $Sr_2RuO_4$ system {\cite{9}.   
From the symmetry point of view $d_{xy}$ does not mix with $d_{yz}$ and
$d_{zx}$, so the fluctuations can be ascribed \cite{10} as solely due to
the electrons in $d_{xy}$ band.  
The Hund's rule coupling between $d_{xy}$ and the remaining two bands $(d_{yz},
d_{zx})$ must than suppress the formation of the antiferromagnetic state.  
In effect, we are left with two electronic liquids: the doubly degenerate
and hybridized 
$d_{yz}-d_{zx}$ band containing approximately one hole and the $d_{xy}$  band
containing the other.  
It must be underlined that all $t_{2g}$ holes are delocalized, since one
observes a well defined Baber-Landau-Pomeranchuk $(\sim T^2)$ contribution to
the resistivity in both $x-y$ $(RuO_2)$ plane and in c 
direction \cite{11} .  

From what has been said above it is important to formulate first the model of
local pairing represented a 
doubly degenerate (or almost degenerate) band coupled by the Hund's rule and
characterize the possible spin-triplet solutions induced by the Hund's rule
(ferromagnetic) exchange.  
This type of model has been formulated by us recently \cite{12}. 
We have shown there that  sizeable (of the order of bare bandwidth) Coulomb
correlations renormalize the system properties, i.e. lead to an {\it almost
localized Fermi liquid} with a nonretarded real-space and spin-triplet 
pairing.  
The renormalized Fermi-liquid nature of our fermionic system will be a 
starting point in this paper, 
in which we consider basic features of the superconducting state such as the
quasiparticle wave function (in the Fock space) and their energies.  
We list the possible solutions for our effective model with interorbital pairing.    
The question of coexistence of the A1 state with ferromagnetism, as well as the
competition with the orbitally ordered-spin ferromagnetic state has been
discussed separately \cite{13}.   
We believe that the present two-band model stands on its own ground,
{\it independently} of the detailed nature of $Sr_2RuO_4$ superconductivity
(which should include the third band and the anisotropic interband 
hybridization) and must be considered separately, to 
amplify the 
physical plausibility of this mechanism of spin-triplet pairing (see also 
the discussion at the end).  
This is particularly so because the Hund's rule and associated with it
ferromagnetic fluctuations \cite{14} represent  probably the most natural
determinants of spin-triplet pairing under these circumstances.  
Also, the present real-space pairing \cite{13} represents is formally analogous
to the 
spin-singlet pairing \cite{5} and additionally, reflects a fundamental
symmetry - the particle interchange.  
So, it contains  fundamental physics   
in  the sense, that the nature of the ground state, i.e. that of the 
spin-triplet superconductor, can 
appear instead of or together with an itinerant ferromagnetism. 

\vskip1.0cm 

\noindent {\Large {\bf II. Nambu-De Gennes method for the triplet pairing  in
the two-band case }}

\vskip0.5cm 

We consider a degenerate two-band Fermi-liquid system coupled by a local triplet
pairing. 
The corresponding effective Hamiltonian is of the simple form \cite{12}
\be 
{\cal{H}} = \sum_{\k  \s  l=1,2} E_{\k l} a_{\k  l \s}^{\dagger}a_{\k  l \s} 
- 2 \tilde{J} \sum_{i m}  A_{im}^{\dagger}A_{im} , 
\label{1}
\ee 
where $E_{\k l}$ are the quasiparticle energies with enhanced masses by the
band narrowing factor $q^{-1}$ (calculated self-consistently \cite{12}) in
the bands $l=1,2$, $\tilde{J} \sim Jt^2$ is the effective Hund's rule coupling
(the local interorbital exchange), and $t^2$ is the probability of having
interorbital local spin-triplet configurations, characterized by the creation
operators $A_{1}^{\dagger} = a_{il\uparrow}^{\dagger} 
a_{il'\uparrow}^{\dagger}$, $A_{-1}^{\dagger} = a_{il\downarrow}^{\dagger}
a_{il'\downarrow}^{\dagger}$, and $A_{0}^{\dagger} = \frac{1}{\sqrt{2}} ( 
a_{il\uparrow}^{\dagger} a_{il'\downarrow}^{\dagger}  + 
a_{il\downarrow}^{\dagger} a_{il'\uparrow}^{\dagger})$ for $l \neq l'$.  
The local exchange origin of the second term derives from the exact relation
between the pairing operators in real space and the full exchange operator
projecting the corresponding two-particle state onto the spin-triplet
configuration. 
\be 
\sum_{m=-1}^{1} A_{im}^{\dagger} A_{im} = {\bf S}_{il} \cdot {\bf S}_{il'} +
\frac{3}{4} n_{il} n_{il'} , 
\label{1a} 
\ee 
where ${\bf S}_{il}$ and $n_{il}$ are respectively the spin and the particle
number operators for electron on site $i$ and orbital $l$.  
Explicitly $n_{il} = \sum_{\s} n_{i l \s}$, $n_{i l \s} = a_{il\s}^{\dagger}
a_{il\s}$, whereas the spin operators ${\bf S}_{il} \equiv (S_{il}^{+},
S_{il}^{-}, S_{il}^{z}) \equiv (a_{il\ua}^{\dagger}a_{il\da}, 
a_{il\da}^{\dagger} a_{il\ua}, (1/2)(n_{il\ua}-n_{il\da}))$.  
The right-hand side of (\ref{2}) represents thus the full exchange operator.  

After making  the BCS-type approximation in the local form \cite{14} 
\be 
A_{im}^{\dagger} A_{im} \simeq A_{im}^{\dagger} <A_{im}> + <A_{im}^{\dagger}>
A_{im} - <A_{im}^{\dagger}><A_{im}> 
\label{2} 
\ee 
we can cast Hamiltonian (\ref{1}) into the four-component form, which in
the reciprocal $(\k )$ space takes the form \cite{12} 
\be 
{\cal{H}}_{BCS} = \sum_{\k} {\bf f}_{\k}^{\dagger} {\bf H}_{\k} {\bf f}_{\k} + 
\sum_{\k} E_{\k 2} , 
\label{3}
\ee 
where the corresponding Nambu operators take the form: 
${\bf f}_{\k}^{\dagger} = ( f_{\k 1 \ua}^{\dagger}, f_{\k 1 \da}, 
f_{-\k 2 \ua}, f_{-\k 2 \da} )$, ${\bf f}_{\k} =
({\bf f}_{\k}^{\dagger})^{\dagger}$, and the Hamiltonian matrix for selected
${\k}$ state reads 
\be 
{\bf H}_{\k} = \left( 
\begin{array}{cccc} 
E_{\k 1} - \mu , & 0, & \Delta_1 , & \Delta_0  \\ 
0 , & E_{\k 1} - \mu , & \Delta_0 , & \Delta_{-1}  \\ 
\Delta_1^{\ast} , & \Delta_0^{\ast} , & -E_{\k 2} + \mu , & 0 \\ 
\Delta_0^{\ast} , & \Delta_{-1}^{\ast} , & 0 , & -E_{\k 2} + \mu 
\end{array} 
\right) 
\equiv 
\left( 
\begin{array}{cc} 
E_{\k 1} \hat{\s}_0 , & \hat{\Delta} \\
\hat{\Delta}^{\ast} , &  -E_{\k 2} \hat{\s}_0  
\end{array} 
\right) , 
\label{4} 
\ee 
where $\hat{\s}_0 \equiv {\bf 1}$ is the unit $2 \times 2$ matrix, and $\mu$ is the
chemical potential. 
The superconducting gap is parametrized as 
$\Delta_m \equiv -2\tilde{J} \sum_{\k} <f_{\k 1 \s}^{\dagger} f_{-\k 2 \s
'}^{\dagger}>$, with $m=(\s + \s ' )/2$, and $\s , \s ' = \pm 1$.   
The $2 \times 2$ matrix $\hat{\Delta}$ is parametrized in the usual
form \cite{16}  
\be 
\hat{\Delta} = i ({\bf d} \cdot \tilde{\s} ) \s_y = 
\left( 
\begin{array}{cc} 
-d_x+id_y, & d_z \\
d_z , & d_x + id_y 
\end{array} 
\right) , 
\label{5} 
\ee 
where $\tilde{\s}$ is composed of the three Pauli matrices, whereas the 
vector ${\bf d}$ in
spin space   has the components 
$d_x = (\Delta_{-1} -
\Delta_1 )/2$, $d_y = (\Delta_{-1} + \Delta_1 )/2$, and $d_z = \Delta_0$. 
The form (\ref{4}) is a generalization of the Nambu representation to the
triplet case with three, in general different, gaps $\Delta_m$.  

It is straightforward to introduce the $4 \times 4$ Dirac matrices 
$$ 
\tilde{\beta} \equiv 
\left( 
\begin{array}{cc} 
\bf{1} , & 0 \\ 
0 , & -\bf{1} 
\end{array} 
\right) 
~~~{\rm and} ~~~ 
\tilde{\alpha}_i = 
\left( 
\begin{array}{cc} 
0 , & \s_i \\
\s_i , & 0 
\end{array} 
\right) , $$ and then rewrite (\ref{4}) for the degenerate case 
$E_{\k 1} = E_{\k 2}$ and for
$\Delta_m = \Delta_m^{\ast}$ in the form 
\be 
{\bf H}_{\k} = \tilde{\beta} ( E_{\k} - \mu ) + i({\bf d} \cdot 
\tilde{\alpha} )
\Sigma_2 , 
\label{6}
\ee 
where $$\Sigma_2 = 
\left( 
\begin{array}{cc} 
\bf{0} , & \s_y \\
\s_y , & \bf{0} 
\end{array} 
\right) $$ is the $y$ component of the relativistic spin operator.  
We discuss in detail the simple situation of degenerate electrons $(E_{\k 1} =
E_{\k 2})$ with a real gap $\Delta_m$ in the next Section.  

One can also look at the approach from a different prospective.  
Let us introduce the four component wave function for a single quasiparticle in
the suprconducting phase propagating 
in the real space as follows 
\be 
\hat{\Psi} ({\bf x} , t ) = \frac{1}{\sqrt{N}} \sum_{\k} 
\left( 
\begin{array}{c} 
\psi_{1\k}f_{\k 1 \ua} \\
\psi_{2\k}f_{\k 1 \da} \\ 
\psi_{3\k}f_{-\k 2 \ua}^{\dagger} \\ 
\psi_{4\k}f_{-\k 2 \da}^{\dagger} 
\end{array}
\right) 
\exp \left[ i\left( \k \cdot {\bf x} - \frac{E_{\k}}{\hbar} t \right) \right] , 
\label{7} 
\ee 
where $\psi_{\mu \k}$ are the quasiparticle amplitudes which are determined for
each eigenstate (see below).  
In this representation the Bogolyubov-De Gennes equation  for a single
quasiparticle in the superconducting states reads:  
\be 
i\hbar \partial_t \hat{\Psi} = \tilde{\beta} \{ E_{\k} ( \k \Rightarrow
\frac{\nabla}{i} ) - \mu \} \hat{\Psi} + i ({\bf d} \cdot \tilde{\alpha} )
\Sigma_2
\hat{\Psi} , 
\label{8}
\ee 
where $E_{\k} (\k \Rightarrow \frac{\nabla}{i})$ represents now the
differential operator $(1/i)\nabla$ replacing the wave vector $\k$ in the
dispersion relation $E_{\k}$ for quasiparticles.  
In the effective-mass approximation and in the stationary case 
this wave equation for quasiparticles in the superconducting phase 
has the following form 
\be 
\lambda \left( 
\begin{array}{c} 
\psi_1 \\
\psi_2 \\
\psi_3 \\
\psi_4 
\end{array} 
\right)  = - \left( \frac{\hbar^2}{2m^{\ast}} \nabla^2 + \mu \right) 
\left( 
\begin{array}{c} 
\psi_1 \\
\psi_2 \\
-\psi_3 \\
-\psi_4 
\end{array} 
\right)  + 
\left( 
\begin{array}{c} 
\Delta_1 \psi_3 + \Delta_0 \psi_4 \\ 
\Delta_0 \psi_3 + \Delta_{-1} \psi_4 \\ 
\Delta_1 \psi_1 + \Delta_0 \psi_2 \\  
\Delta_0 \psi_1 + \Delta_{-1} \psi_2 
\end{array} 
\right) , 
\label{9} 
\ee 
where $\psi_{\mu} \equiv \psi_{\mu}({\bf x})$ and 
$\lambda$ is an eigenvalue of quasiparticle state in the superconducting
state with the above 4-component wave function ($\Delta_m$ are regarded as
real).    
The validity of this equation goes beyond the simple solution (\ref{7}), as one
can include the magnetic and electric fields and other inhomogeneities if they
appear on the mesoscopic or macroscopic scale.  
In the next Section we will use explicitly the momentum representation of
Eqs.(\ref{9}), as we will discuss exclusively homogeneous superconducting
states.  
We will return to Eqs.(\ref{9}) when discussing the general features of this
Hamiltonian in Section IV. 
One should also note that finding the eigenvalues for Hamiltonian  in the forms
(\ref{3}) or (\ref{6}) can be achieved by diagonalizing of the matrix $4
\times 4$ in general case, as discussed in analytic terms in Appendix A.  

\vskip1.5cm

\noindent {\Large  {\bf III. Superconducting states and their quasiparticles} }

\vskip0.5cm

We now discuss three principal solutions of Eq.(\ref{9}) by taking $\psi_{\mu}
( {\bf x}) = \psi_{\mu} \exp (i \k \cdot {\bf x} ) /\sqrt{V}$, where $V$ is the
system volume.  
We also assume that $\Delta_{\mu} = \Delta_{\mu}^{\ast}$, (e.g. neglect the
applied magnetic fields), since we consider only spatially homogeneous 
solutions.  
Namely, rewriting Eq.(\ref{9}) in components we obtain the combinations 
\be 
\left\{ 
\begin{array}{c}
 \lambda (\psi_1 + \psi_2 ) = (E_{\k}-\mu ) (\psi_1 + \psi_2)+(\Delta_1 +
\Delta_0)\psi_3 + (\Delta_0+\Delta_{-1})\psi_4 \\ 
\lambda (\psi_3 + \psi_4 ) = -(E_{\k}-\mu ) (\psi_3 + \psi_4)+(\Delta_1 +
\Delta_0)\psi_1 + (\Delta_0+\Delta_{-1})\psi_2  ,
\end{array} 
\right. 
\label{10} 
\ee   
and 
\be 
\left\{ 
\begin{array}{c}
 \lambda (\psi_1 - \psi_2 ) = (E_{\k}-\mu ) (\psi_1 - \psi_2)+(\Delta_1 -
\Delta_0)\psi_3 + (\Delta_0-\Delta_{-1})\psi_4 \\ 
\lambda (\psi_3 - \psi_4 ) = -(E_{\k}-\mu ) (\psi_3 - \psi_4)+(\Delta_1 -
\Delta_0)\psi_1 + (\Delta_0-\Delta_{-1})\psi_2  .
\end{array} 
\right.  
\label{11} 
\ee  
Such combinations of particle $(\psi_1$ and $\psi_2)$ and hole ($\psi_3$ and
$\psi_4$) components contain basic symmetry, as we will see on example of
particular solutions, which we discuss next.  

\vskip0.5cm 

\noindent {\large {\bf A. Isotropic solution: $\Delta_0 = \Delta_{-1} = \Delta_1
\equiv \Delta$ }}

\vskip0.5cm 
 
In that situation Eqs.(\ref{10}) - (\ref{11}) take a simple form 

\be 
\left\{ 
\begin{array}{c}
 \lambda (\psi_1 + \psi_2 ) = (E_{\k}-\mu ) (\psi_1 + \psi_2)+
 2\Delta (\psi_3 + \psi_4 ) \\ 
\lambda (\psi_3 + \psi_4 ) = -(E_{\k}-\mu ) (\psi_3 + \psi_4)+
2\Delta (\psi_1 + \psi_2 )   ,
\end{array} 
\right.  
\label{12} 
\ee  
and 
\be 
\left\{ 
\begin{array}{c}
 \lambda (\psi_1 - \psi_2 ) = (E_{\k}-\mu ) (\psi_1 - \psi_2)  \\ 
 \lambda (\psi_3 - \psi_4 ) = -(E_{\k}-\mu ) (\psi_3 - \psi_4)  . 
\end{array} 
\right.  
\label{13} 
\ee   
The first two equations lead to the modes with a gap 
\be 
\lambda = \lambda_{\k 1,2} = \pm \sqrt{ (E_{\k} - \mu )^2 + 4 \Delta^2} \equiv
\pm \lambda_{\k} . 
\label{14} 
\ee 
For those two modes $\psi_1 = \psi_2$ and $\psi_3 = \psi_4$ and their
eigenstates are characterized by the following quasiparticle operators 
\be 
\alpha_{\k} = u_{\k} \frac{1}{\sqrt{2}} \left( f_{\k 1\ua} + f_{\k 1\da}
\right) - v_{\k} \frac{1}{\sqrt{2}} \left( f_{-\k 2\ua}^{\dagger} 
+ f_{-\k 2\da} \right) , 
\label{15} 
\ee 
and 
\be 
\beta_{-\k}^{\dagger} = v_{\k} \frac{1}{\sqrt{2}} \left( f_{\k 1\ua} + f_{\k 1\da}
\right) + u_{\k} \frac{1}{\sqrt{2}} \left( f_{-\k 2\ua}^{\dagger} 
+ f_{-\k 2\da} \right) , 
\label{16} 
\ee  
with the Bogolyubov coherence factors 
\be 
u_{\k} = \frac{1}{\sqrt{2}} \left( 1 + \frac{E_{\k} - \mu}{\lambda_{\k}}
\right)^{1/2} , ~~~~~
v_{\k} = \frac{1}{\sqrt{2}} \left( 1 - \frac{E_{\k} - \mu}{\lambda_{\k}}
\right)^{1/2} . 
\label{17} 
\ee 
The quasiparticle operators  contain symmetric combinations $(f_{\k 1\ua} +
f_{\k 1 \da})/\sqrt{2}$  and $(f_{-\k 2\ua} +
f_{-\k 2 \da})/\sqrt{2}$. 
The wave function is symmetric with respect to particle-spin interchange $(\ua
\leftrightarrow \da )$  and describes  quasiparticle 
states of energy $\pm \lambda_{\k}$, respectively.  

Eqs.(\ref{13}) lead to the gapless modes of the form 
\be 
\lambda = \lambda_{\k 3,4} = \pm (E_{\k} - \mu )  , 
\label{18} 
\ee 
and correspond to the eigenstates characterized by the operators 
\be 
\gamma_{\k} = \frac{1}{\sqrt{2}} \left( f_{\k 1 \ua} - f_{\k 1 \da} \right) ,
~~~{\rm and}~~~~
\delta_{-\k}^{\dagger} = \frac{1}{\sqrt{2}} \left( f_{-\k 2 \ua}^{\dagger} -
f_{-\k 2 \da}^{\dagger} \right) 
\label{19} 
\ee 
and constitute the antisymmetric-in-spin operators, representing the unpaired
electrons.  
These gapless modes disappear when the gap components are not equal, as shown
in Appendix A. 
One should note that the gapless modes appear even though the superconducting
gap here is $\k$-independent. 

Combining the solutions (\ref{15}) - (\ref{17}) and (\ref{18} - (\ref{19}) we
can express the original ("old") particle operators in terms of quasiparticle
("new") operators in the following manner 
\be 
\left( 
\begin{array}{c} 
f_{\k 1 \ua} \\ 
f_{\k 1 \da} \\ 
f_{-\k 2 \ua}^{\dagger} \\ 
f_{-\k 2 \da}^{\dagger} 
\end{array} 
\right) 
= \frac{1}{\sqrt{2}} 
\left( 
\begin{array}{cccc} 
u_{\k} , & v_{\k} , & 1 , & 0 \\ 
u_{\k} , & v_{\k} , & -1 , & 0 \\
-v_{\k} , & u_{\k} , & 0 , & 1 \\ 
-v_{\k} , & u_{\k} , & 0 , & -1 
\end{array} 
\right) 
\left( 
\begin{array}{c} 
\alpha_{\k} \\ 
\beta_{-\k}^{\dagger} \\ 
\gamma_{\k} \\ 
\delta_{-\k}^{\dagger} 
\end{array} 
\right) . 
\label{20} 
\ee 
This transformation is  necessary for determining the self-consistent equation
for the gap and for the chemical potential $\mu$.  
First, we rewrite the Hamiltonian (\ref{3}) in the diagonal form 
$$
{\cal{H}} = \sum_{\k} \lambda_{\k} (\alpha_{\k}^{\dagger} \alpha_{\k} -  
\beta_{-\k} \beta_{-\k}^{\dagger}  ) + E_{\k} 
 (\gamma_{\k}^{\dagger} \gamma_{\k} -  
\delta_{-\k} \delta_{-\k}^{\dagger}  ) + \sum_{\k} E_{\k} $$
\be 
=  \sum_{\k} \lambda_{\k} (\alpha_{\k}^{\dagger} \alpha_{\k} + 
\beta_{-\k}^{\dagger} \beta_{-\k} - 1  ) + E_{\k} 
 (\gamma_{\k}^{\dagger} \gamma_{\k} +  
\delta_{-\k}^{\dagger} \delta_{-\k} )  . 
\label{21} 
\ee 
The equation for the gap e.g. $\Delta_1 = < f_{\k 1 \ua}^{\dagger} f_{-\k
2\ua}^{\dagger}>$ is obtained by substituting the relevant transformed 
operators in   
(\ref{20}) to $\Delta_1$.  
In effect, we obtain the usual BCS form $(E_{\k} \equiv E_{\k} - \mu )$ 
\be 
<f_{\k 1 \ua}^{\dagger}f_{-\k2\ua}^{\dagger} > = - \frac{1}{2}
\frac{\Delta}{\sqrt{E_{\k}^2+4\Delta^2}} \tanh \left( \frac{\beta
\sqrt{E_{\k}^2+4\Delta^2}}{2} \right) , 
\label{22} 
\ee 
where $\beta \equiv (k_BT)^{-1}$. So, the gap equation  has two solutions: 
$ 
1^o ~~ \Delta \equiv 0 ,$

\noindent $2^o$ 
\be  
1 = \frac{J}{N} \sum_{\k} \frac{1}{\sqrt{E_{\k}^2 + 4\Delta^2}} \tanh
\left( \frac{\beta \sqrt{E_{\k}^2+4\Delta^2}}{2} \right) . 
\label{23} 
\ee 
The last equation tells us that the physical gap is $2\Delta$.
The self-consistent equation for the chemical potential must include gapless
modes, i.e. takes the form 
\be 
n = \frac{1}{N} \sum_{\k \s} <f_{\k 1 \s}^{\dagger} f_{\k 1 \s} + f_{\k 2
\s}^{\dagger} f_{\k 2\s} > = \frac{2}{N} \sum_{\k} <\alpha_{\k}^{\dagger}
\alpha_{\k} + \gamma_{\k}^{\dagger} \gamma_{\k} > . 
\label{24}
\ee 
Normally, as we shall see, $\Delta \ll | \mu |$, and hence approximately half of
all particles will have the spectrum gapped.  
The details must be analysed numerically for a concrete structure of the
density of states.  
In the limit $\tilde{W} \ll \tilde{J}$ we have the estimate of the gap 
value  
 at $T=0$ in the form $\Delta = (\tilde{W}/2)\exp (-\tilde{W}/(2\tilde{J} )$; 
this yields the value $\Delta /\tilde{W} \sim 10^{-3}-10^{-4}$, or in the 
regime 
$1-10K$ for $\tilde{W} \simeq 1eV$ and $\tilde{J} \sim 0.1\tilde{W}$.  

We need also the expression for the ground state energy, as various solutions
are possible.  
In the present case, this energy can be written as 
\be 
\frac{E_G}{N} = \frac{2}{N} \sum_{\k} \{ \sqrt{E_{\k}^2+ \tilde{\Delta}^2}<
\alpha_{\k}^{\dagger}\alpha_{\k} > + E_{\k} <\gamma_{\k}^{\dagger} \gamma_{\k}>
- \sqrt{E_{\k}^2+\tilde{\Delta}^2} \} + \frac{\tilde{\Delta}^2}{2J} , 
\label{24a} 
\ee 
where $\tilde{\Delta} = 2 \Delta$ .

\vskip1.0cm

\noindent {\large {\bf B. Equal-spin pairing: $\Delta_0 \equiv 0$ }}

\vskip0.5cm 

To obtain the explicit solution we now combine separately the first and third
components of Eq.(\ref{9}) on one side, and the second and the fourth on 
the other. 
Adding and subtracting the corresponding terms we obtain: 
\be 
\left\{ 
\begin{array}{c} 
(\Delta_1 - \lambda )(\psi_1 + \psi_3 ) + (E_{\k}- \mu ) (\psi_1 - \psi_3 ) = 0
\\ 
(E_{\k} - \mu ) (\psi_1 + \psi_3) - (\Delta_1 + \lambda )(\psi_1 - \psi_3) = 0
, 
\end{array} 
\right.  
\label{25} 
\ee 
and 
\be 
\left\{ 
\begin{array}{c} 
(\Delta_{-1} - \lambda )(\psi_2 + \psi_4 ) + (E_{\k}- \mu ) (\psi_2 - 
\psi_4 ) = 0 \\  
(E_{\k} - \mu ) (\psi_2 + \psi_4) - (\Delta_{-1} + \lambda )(\psi_2 - 
\psi_4) = 0  . 
\end{array} 
\right.  
\label{26} 
\ee 
Thus, the two pairs of components (\ref{25}) and (\ref{26}) separate from each
other and it is sufficient to solve e.g. the first system (\ref{25}) to be able
to reproduce the other. 
Explicitly, the two solutions can be combined into the form, in which the
eigenvalues take the form 
\be 
\lambda \equiv \lambda_{\k 1...4} = \pm \sqrt{(E_{\k}-\mu )^2 + \Delta_{\s}^2}
\equiv \pm \lambda_{\k}^{(\s )} , 
\label{27} 
\ee 
where for each spin orientation $\s = \pm 1$ of the quasiparticles we have two
solutions with the gap $\pm \sqrt{(E_{\k}-\mu )^2 + \Delta_{\s}^2}$. 
The quasiparticle operators $(\alpha_{\k \s}, \beta_{-\k \s}^{\dagger})$
diagonalizing Hamiltonian (\ref{3}) in this case are: 
\be 
\alpha_{\k \s} = u_{\k}^{(\s )} \frac{1}{\sqrt{2}} \left( f_{\k 1\s} + 
f_{-\k 2\s}^{\dagger}  
\right) - v_{\k}^{(\s )} \frac{1}{\sqrt{2}} \left( f_{\k 1\s}  
- f_{-\k 2\s}^{\dagger} \right) , 
\label{28} 
\ee 
and 
\be 
\beta_{-\k \s}^{\dagger} = -v_{\k}^{(\s )} \frac{1}{\sqrt{2}} 
\left( f_{\k 1\s} + f_{-\k 2\s}^{\dagger} 
\right) + u_{\k}^{(\s )} \frac{1}{\sqrt{2}} \left( f_{\k 1\s}  
- f_{-\k 2\s}^{\dagger} \right) , 
\label{29} 
\ee  
with the coherence factors 
\be 
u_{\k}^{(\s )} = \frac{1}{\sqrt{2}} \left( 1 + \frac{\Delta_{\s
}}{\lambda_{\k}^{(\s )}} 
\right)^{1/2} , ~~~~~
v_{\k}^{(\s )} = \frac{1}{\sqrt{2}} \left( 1 - \frac{\Delta_{\s
}}{\lambda_{\k}^{(\s )}}
\right)^{1/2} . 
\label{29a} 
\ee 
In general, we have two gaps $\Delta_{\s} = (\Delta_1 , \Delta_{-1} )$.  
In the situation $\Delta_{\s} = \Delta_{-\s} = \Delta$ we have a doubly (spin)
degenerate solutions.  
It can be easily verified that the operators (\ref{28}) and (\ref{29}) obey the
fermion anticommutation relations.  
The diagonalized Hamiltonian has the form 
\be 
{\cal{H}} = \sum_{\k} \lambda_{\k}^{(\s )} \left( \alpha_{\k \s}^{\dagger}
\alpha_{\k \s} + \beta_{\k \s}^{\dagger} \beta_{\k \s} - 1 \right) + E_0 . 
\label{29b} 
\ee 

To determine the gap equation we have to find the transformation which is
reverse of (\ref{28}) and (\ref{29}).  
It is of the form 
\be 
\left( 
\begin{array}{c} 
f_{\k 1 \ua} \\ 
f_{\k 1 \da} \\ 
f_{-\k 2 \ua}^{\dagger} \\ 
f_{-\k 2 \da}^{\dagger} 
\end{array} 
\right) 
= \frac{1}{\sqrt{2}} 
\left( 
\begin{array}{cccc} 
u_{\k}^{(+)}+v_{\k}^{(+)} , & u_{\k}^{(+)}-v_{\k}^{(+)} , & 0 , & 0 \\ 
0 , & 0 , & u_{\k}^{(-)}+v_{\k}^{(-)} , & u_{\k}^{(-)}-v_{\k}^{(-)} \\
u_{\k}^{(+)}-v_{\k}^{(+)} , & -u_{\k}^{(+)}-v_{\k}^{(+)} , & 0 , & 0 \\ 
0 , & 0 , & u_{\k}^{(-)}-v_{\k}^{(-)} , & -u_{\k}^{(-)}-v_{\k}^{(-)} 
\end{array} 
\right) 
\left( 
\begin{array}{c} 
\alpha_{\k \ua} \\ 
\beta_{-\k \ua}^{\dagger} \\ 
\alpha_{\k \da} \\ 
\beta_{-\k \da}^{\dagger} 
\end{array} 
\right) . 
\label{30} 
\ee 
The difference with the isotropic pairing (\ref{20}) is that here the coherence
factors appear in combinations.  
Those appear also in the self-consistent equation for the gap 
\be 
<f_{-\k2\s}^{\dagger}f_{\k 1 \s}^{\dagger}> = -(u_{\k}^{(\s )2}-v_{\k}^{(\s
)2})[1-<\alpha_{\k\s}^{\dagger}\alpha_{\k \s}> - <\beta_{\k
\s}^{\dagger}\beta_{\k \s}>] . 
\label{31} 
\ee 
In result, the self-consistent equation will have the following three 
solutions: 
\noindent $1^o$ $\Delta_{\s } \equiv 0$ ,
\noindent $2^0$ $\Delta_{(-\s )} = 0$, but $\Delta_{\s } \neq 0$ is the
solution of equation 
\be 
1 = \frac{J}{N} \sum_{\k} \frac{1}{\sqrt{(E_{\k}-\mu )^2 + \Delta_{\s }^2}} 
\tanh
\left( \frac{\beta}{2} \sqrt{(E_{\k}-\mu )^2+\Delta_{\s }^2} \right) 
\label{32} 
\ee 
\noindent $3^o$ $\Delta_{\s } \neq 0$, $\Delta_{-\s} \neq 0$, and each of them
is determined from equation (\ref{32}). 

One should note that the coupling constant above $(J)$ is the same as for the
isotropic phase (cf. Eq.(\ref{23})). 
For the sake of completness, we reproduce the ground-state-energy expression
which is 
$$
\frac{E_G}{N} = \frac{2}{N} \sum_{\k \s} \sqrt{(E_{\k} - \mu )^2 +
\Delta_{\s}^2} <\alpha_{\k \s}^{\dagger} \alpha_{\k \s} > +
\frac{\Delta_1^2+\Delta_{-1}^2}{2J} $$ 
\be 
- \frac{1}{N} \sum_{\k \s} \sqrt{(E_{\s} -
\mu )^2 + \Delta_{\s}^2} + \frac{1}{N} \sum_{\k} E_{\k} . 
\label{32a} 
\ee 
This phase represents a starting point when discussing the coexistence of 
ferromagmnetism and the spin-triplet superconductivity.

\vskip1.0cm

\noindent {\large {\bf C. Spin-polarized phase: $\Delta_0 = \Delta_{\da} = 0$  
}}

\vskip0.5cm 

In this limit the system is totally spin polarized, i.e. is a {\it 
spin-saturated  superconductor}.  
In that limit we recover again  the spectrum both with and without gap, i.e.
$\lambda_{\k 1,2} = \pm \sqrt{(E_{\k} - \mu )^2 + \Delta_{\ua}^2}$,
$\lambda_{\k 3,4} = \pm (E_{\k} - \mu )$.  
Thus paired and unpaired states coexist also in this phase, 
as can be easily seen from 
Eqs.(\ref{28}) - (\ref{29}), which yield the form written there for $\s = \ua$
and $\alpha_{\k \da} = f_{\k 1 \da}$ and $\beta_{-\k \da}^{\dagger} = -f_{\k 2
\da}^{\dagger}$.  

\vskip0.5cm 

Summarizing cases A-C, the lowest energy will have the homogeneous state with
$\Delta_{\ua} = \Delta_{\da} = \Delta_0$ so that the effective gap is equal to
$2\Delta$.  
The most interesting feature of the results is that the gapless modes coexist
in cases A and C and represent half of the spectrum.  
Also, the condensed phases described by the 
 cases A-C above correspond roughly to the solutions for
superfluid $^3He$, which are labelled B, A, and A1.  
However, under the present circumstances here we have momentum {\it
independent} gaps, since the pairing is of the local (intrasite, but 
interorbital) nature.  
For the sake of comparison we present in Appendix B the case of spin-singlet
pairing induced by the same type of local interband pairing induced by 
antiferromagnetic exchange.  

\vskip1.5cm

\noindent {\Large  {\bf IV. Remark on the triplet pairing for relativistic 
fermions } } 

\vskip0.5cm

The two-band situation with a local ferromagnetic exchange can be easily
generalized to the explicitly relativistic form modelling thus the triplet
configuration of spin, isospin or color (the singlet case was considered by
Nambu and Jona-Lasinio \cite{16} and in  \cite{19}).  
The paired quasiparticles \cite{20} obey the following modified Dirac wave
equation 
\be 
i\hbar \partial_t \Psi =  (c \tilde{\alpha} \cdot \hat{\bf p} + \tilde{\beta} mc^2
)\Psi + i ({\bf d} \cdot \tilde{\alpha}) \Sigma_2 \Psi , 
\label{38}  
\ee 
where the last term supplements the Dirac equation with the contact pairing. 
By taking the analogy with the original approach by Nambu \cite{21} one can
write down the stationary version of this equation as the following system in
the two-component (Weyl) representation 
\be  
\lambda 
\left(  
\begin{array}{c} 
\psi_1 \\
\psi_2 
\end{array}
\right) 
= (c \tilde{\s} \cdot \hat{\bf p} - \mu ) 
\left(  
\begin{array}{c} 
\psi_1 \\
\psi_2 
\end{array}
\right) 
+mc^2 
\left(  
\begin{array}{c} 
\psi_3 \\
\psi_4 
\end{array}
\right) + 
\left(  
\begin{array}{c} 
\Delta_1 \psi_3 + \Delta_0 \psi_4 \\
\Delta_0\psi_3 + \Delta_{-1}\psi_4  
\end{array}
\right) ,
\label{39}
\ee 

\be  
\lambda 
\left(  
\begin{array}{c} 
\psi_3 \\
\psi_4 
\end{array}
\right) = 
-(c \tilde{\s} \cdot \hat{\bf p} - \mu ) 
\left(  
\begin{array}{c} 
\psi_3 \\
\psi_4 
\end{array}
\right) 
+mc^2 
\left(  
\begin{array}{c} 
\psi_1 \\
\psi_2 
\end{array}
\right) + 
\left(  
\begin{array}{c} 
\Delta_1 \psi_1 + \Delta_0 \psi_2 \\
\Delta_0\psi_1 + \Delta_{-1}\psi_2  
\end{array}
\right) . 
\label{40}
\ee 
This system of equations can be directly compared with Eq.(\ref{9}) for
nonrelativistic electrons. 
In the present situation the mass term mixes the upper and the lower two
components of the bispinor, as does the pairing part. 
Apart from a modification in the kinetic-energy part, the system of equations
(\ref{39})-(\ref{40}) can be solved in the manner as discussed in Section III. 
The detailed discussion must include also the gauge fields, which lead to
one-exchange pairing (analogous to the phonon-mediated pairing), a topic
intensively discussed in recent literature \cite{22}. 
In general, the singlet pairing \cite{16}, \cite{19} - \cite{22} is 
mutually exclusive with 
the triplet pairing proposed here and therefore, the latter requires first a
detailed discussion of exchange interactions represented by relativistic spin
$\{ \Sigma_i \}$.

\vskip1.0cm

\noindent {\Large  {\bf V. Discussion } }

\vskip0.5cm

In this paper we have formulated the quasiparticle language for local triplet
pairing between fermions (interband pairing in the nonrelativistic case)
induced by the local (Hund's rule or Dirac) exchange.  
In particular, we have determined explicitly the quasiparticle states and the
De Gennes wave equation for them, which can be useful when considering
spatially  inhomogeneous situations \cite{23}.  
The principal feature of the results is the existence of the gapless modes,
existence of which can also be proved on a phenomenological level \cite{24}.  
The circumstance that the pairing is induced by the ferromagnetic exchange
means that this interaction can lead not only to an itinerant ferromagnetic
state, but also to either spin-triplet superconductor or to a coexistence of
both these states (for a brief discussion of this issue see Ref.\cite{13}). 
The present paper represents only the first step in this direction. 
Furthermore, our mechanism of pairing expressing the fundamental symmetry (the
particle interchange) may have an important astrophysical application: the
pairing in the neutron-proton matter in pulsar, but this intriguing possibility
requires a separate study.  

Two problems should be tackled next.  
First, the analysis of the Meissner effect, since in the present situation the
orbital diamagnetism  will compete with the ferromagnetic spin polarization
(particularly, if the triplet superconducting and ferromagnetic phases can
coexist).  
An intriguing question here is: can we reach the limiting superconducting phase
(corresponding to A1 phase in the case of superfluid $^3He$), the critical
temperature $T_c$ of which can be enhanced by the applied magnetic field?  

Second, one should derive microscopically the Ginzburg-Landau equation for the
condensed pairs. 
Note that the De Gennes equation (\ref{8}) or (\ref{9}) is useful in describing
the quasiparticle tunneling, whereas the Ginzburg-Landau equation is useful
when considering the Josephon (pair) tunneling.  
Here an intriguing question to what extent the gapless 
quasiparticles influence the tunneling between the spin-singlet and 
the spin-triplet 
superconductors or between the triplet superconductor and normal metal.  
We should be able to see the progress in answering those questions in a near
future.  

Finally, returning to the question of the nature of the paired state in
$Sr_2RuO_4$ one can make the following two remarks. 
First, the existence of gapless modes in B- and A1-like phases leads to the
persistence of the linear term in the specific heat in the superconduncting
phase at its $50\%$ value, if the pairing is the pure spin-triplet state of
electrons pairs derived from $d_{zx}$ bands. 
The recent measurements \cite{25} in very pure samples contradict such earlier
claims \cite{11} that a half of the linear specific heat survives in the
superconducting phases. 
Does that mean that the full phase diagram involves more than one phase 
depending on 
the doping degree, as in the heavy-fermion system 
$U_{1-x}Th_xBe_{13}$ \cite{26} ? 
In connection with this one can say that because of the reasons mentioned in
Section I 
it is conceivable that a singlet pairing 
in $d_{xy}$ band induced by antiferromagnetic fluctuations \cite{9} can compete
in the triplet state in the other two bands \cite{27}. 
The nature of the resultant state should be determined then. 
We should be able to see a progress in those matters in near future.  

Second, an important question concerns the nature of the pairing potential.  
In more standard approach \cite{28}, \cite{14} one introduces the effective 
triplet pairing via the pa\-ra\-mag\-non exchange. 
In that situation the coupling constant is determined by the susceptibility
$\chi ({\bf q} = \k - \k ')$ and hence, is wave vector (${\bf q}$)  
{\it dependent}.  
In the approach developed here the exchange interaction itself provides real
space pairing, as in the case of high temperature superconductors \cite{5}. 
In the case of Hund's rule coupling the pairing potential is then
$\k$-independent. 
We believe that the latter approach is relevant when the particles are strongly
correlated. 
$Sr_2RuO_4$ is a systems close to (but below) the Mott-Hubbard localization
threshold, i.e. the halfway between the weak-correlation and the 
strong-correlation asymptotic regimes.  
Therefore, the real space pairing  is certainly worth of analysing, as it
allows for an analytic approach.

One methodological remark at the end.  
In the analytical analysis of the spin-triplet pairing one usually uses
\cite{17}, \cite{29} the ${\bf d}$ vector in expressing the pairing part. 
Here we decided to use the original BCS gap parameters, a completely equivalent
procedure,  but probably  a bit more direct, at least in the spatially
homogenous situation.  In connection with this a difference of the present
description with that for superfluid $helium-3$ should be stressed.
Namely, in the $helium-3$ case, the $L=1$ orbital moment ${\bf l}$ and the spin 
vector ${\bf d}$ determine the (many-component) nature of the gap. The ${\bf
 d.l}$ and ${\bf dxl}$ combinations determine the order-parameter dynamics.
Here, no ${\bf l}$ vector appears and therefore, the order parameter can have
up to 3 independent components.

\vskip1.0cm 

\noindent {\Large  {\bf Acknowledgments } } 

\vskip0.5cm

I am grateful to my student Andrzej Klejnberg for the cooperation,
which lead to this effective model. 
I am also grateful to W\l{}odek W\'ojcik for many discussions and a technical
help. 
Additionally, I was partially motivated by the question posed by Mark Jarrell 
from the 
University of Cincinnati, who asked if a simpler model representation of the
work presented in Ref.12 was possible. 
The work was  supported by KBN, Grant No. 2P03B 092 18.

\vskip1.5cm

\newpage

\noindent {\Large  {\bf Appendix A: General solution: $\Delta_1 \neq \Delta_2
\neq \Delta_3 \neq \Delta_1$, ~$E_{\k 1} \neq E_{\k 2}$ } } 

\vskip0.5cm

The most general case of finding the eigenvalues for quasiparticles 
in the superconducting phase
is to diagonalize matrix $4 \times 4$, i.e. to solve the equation (we assume
that $E_{\k 1,2} - \mu \equiv E_{\k}$ and that $\Delta_1 = \Delta_1^{\ast}$) 
\be 
det \left( 
\begin{array}{cccc} 
E_{\k 1} - \lambda , & 0 , & \Delta_1 , & \Delta_0 \\
0 , & E_{\k 1} - \lambda ,  & \Delta_0 , & \Delta_{-1} \\
\Delta_1 , & \Delta_0 , & -E_{\k 2} - \lambda , & 0 \\
\Delta_0 , & \Delta_{-1} , & 0 , & -E_{\k 2} - \lambda  
\end{array} 
\right) = 0 . 
\label{a1} 
\ee 
By a straightforward evaluation one obtains the eigenvalues 
\be 
\lambda_{\k1..4} = \frac{1}{2} (E_{\k 1} - E_{\k 2}) \pm \frac{1}{2} \left[ 
 (E_{\k 1} + E_{\k 2})^2 + 2\tilde{\Delta}^2 \pm 2\sqrt{\tilde{\Delta}^4 - 4
\tilde{\delta}^4} \right]^{1/2} , 
\label{a2} 
\ee 
where 
\be 
\tilde{\Delta} = (\Delta_1^2+ \Delta_{-1}^2 + 2 \Delta_0^2 )^{1/2}, 
\label{a3}
\ee 
is the total gap, and 
\be 
\tilde{\delta} = \sqrt{| \Delta_0^2 - \Delta_1 \Delta_{-1} |} , 
\label{a4} 
\ee 
is its anisotropy in fermion-pair spin space. 
In this general case all four modes are each with a different gap  and the
results reduce nicely to the eigenvalues discussed as cases A-C in  main
text. 
In general, the superconducting coupling at the level of energies amounts to
hybridizing the different fermion fields $(l=1,2)$ and their spin 
$(\s = \ua , \da )$ states.  
The general form of the eigenstates can also be found in a straightforward
manner, but will not be discussed here.

\vskip1.5cm

\newpage 

\noindent {\Large  {\bf Appendix B: Local spin-singlet pairing 
in two-band case } } 

\vskip0.5cm

For the sake of comparison with the spin-triplet case we outline here the
solution for the corresponding spin-singlet situation.  
In this case the Hamiltonian with the spin-singlet exchange coupling has the
form in the real space 
\be 
{\cal{H}} = \sum_{\k \s l=1,2} E_{\k l} n_{\k l \s} + J \sum_{i l l'} ( {\bf
S}_{il} {\bf S}_{il'} + \frac{1}{4} n_{il}n_{il'} ) . 
\label{b1} 
\ee 
Note that now the  exchange operator in the present situation 
differs from (\ref{1a}) introduced 
in the triplet case. 
Introducing the corresponding local pairing operators in the singlet state 
\be 
B_i^{\dagger} = \frac{1}{\sqrt{2}} \left( a_{i1 \ua}^{\dagger} 
a_{i 2 \da}^{\dagger} - a_{i 1 \da}^{\dagger} a_{i2  \ua}^{\dagger} \right) 
\label{b2} 
\ee 
we can write down the second term in (\ref{b1}) as $-2J\sum_{i} B_i^{\dagger}
B_i$.  

After taking the space Fourier transform, and including only $(\k , - \k )$
pairs we obtain 
\be 
{\cal{H}} = \sum_{\k \s} E_{\k l} n_{\k l \s} - \frac{J}{N} \sum_{\k \k '} \left(
f_{\k 1 \ua}^{\dagger}f_{-\k 2 \da}^{\dagger} - f_{\k 1 \da}^{\dagger} 
f_{-\k 2 \ua}^{\dagger} \right) \left( f_{-\k 2 \da} f_{\k 1 \ua} - 
f_{-\k 2 \ua} f_{\k 1 \da} \right) . 
\label{b3} 
\ee 
Making subsequently, as in Section II the BCS approximation, we obtain (the
chemical potential is included in $E_{\k l} \equiv E_{\k l} - \mu $) 
\be 
{\cal{H}}_{BCS} = \sum_{\k l \s} \{ E_{\k l} n_{\k l \s} + \Delta_{\k}^{\ast}
\left( f_{\k 1 \ua}^{\dagger} f_{-\k 2 \da}^{\dagger} - 
f_{\k 1 \da}^{\dagger}f_{-\k 2 \ua}^{\dagger} \right) + H.C. \} - 
\frac{\Delta^2}{2J} N , 
\label{b4} 
\ee 
where 
\be 
\Delta \equiv - \frac{2J}{N} \sum_{\k} < f_{\k 1 \ua}^{\dagger} 
f_{-\k 2 \da}^{\dagger}  > . 
\label{b5}
\ee 
Introducing, as before, the four dimensional vectors 
${\bf f}^{\dagger} \equiv (f_{\k 1 \ua}^{\dagger} , f_{\k 1 \da}^{\dagger} , 
f_{-\k 2 \ua}  , f_{-\k 2 \da} )$ and their conjugate as one column vectors, we
can write down the Hamiltonian in the form of the following $4 \times 4$ 
matrix 
\be 
{\cal{H}}_{BCS} = E_0 + 
\left( 
\begin{array}{cccc} 
f_{\k 1 \ua}^{\dagger} , & f_{\k 1 \da}^{\dagger} , &  
f_{-\k 2 \ua}  ,  & f_{-\k 2 \da} 
\end{array} 
\right) 
\left( 
\begin{array}{cccc} 
E_{\k 1} , &  0 , &  0 ,  &  \Delta  \\
0 , & E_{\k 1} , & - \Delta , &  0 \\ 
0 , & - \Delta^{\ast} , & - E_{\k 2} , & 0 \\
\Delta^{\ast} , & 0 , & 0 , & -E_{\k 2} 
\end{array} 
\right) 
\left( 
\begin{array}{c} 
f_{\k 1 \ua} \\ 
f_{\k 1 \da} \\ 
f_{-\k 2 \ua}^{\dagger} \\ 
f_{-\k 2 \da}^{\dagger}  
\end{array} 
\right) 
\label{b6}
\ee 
with $E_0 = 2\sum_{\k} E_{\k 2} + \Delta^2/(2J)$.  
Diagonalization of this $4 \times 4$ matrix leads to the eigenvalues 
\be 
\lambda_{1,2} = \frac{1}{2} \left( E_{\k 1} - E_{\k 2} \right) \pm 
\sqrt{ \left( \frac{E_{\k 1} + E_{\k 2}}{2} \right)^2 + | \Delta |^2 } . 
\label{b7} 
\ee 
Both eigen modes are doubly degenerate and with an isotropic gap $\Delta$. 
We take the form of usual dispersion relation for degenerate bands $E_{\k 1} =
E_{\k 2}$.  

The corresponding combinations of the wave function components are (for $\Delta
= \Delta^{\ast}$) 
\be 
\left\{ 
\begin {array}{c} 
(E_{k 1} - \lambda ) (\psi_1 + \psi_2 ) - \Delta (\psi_3 - \psi_4 ) = 0 \\ 
\Delta (\psi_1 + \psi_2 ) + (E_{\k 2} + \lambda )(\psi_3 - \psi_4 ) = 0  , 
\end{array} 
\right. 
\label{b8} 
\ee  
and 
\be 
\left\{ 
\begin {array}{c} 
(E_{k 1} - \lambda ) (\psi_1 - \psi_2 ) + \Delta (\psi_3 + \psi_4 ) = 0 \\ 
\Delta (\psi_1 - \psi_2 ) - (E_{\k 2} + \lambda )(\psi_3 + \psi_4 ) = 0  . 
\end{array} 
\right. 
\label{b9} 
\ee 
For the sake of simplicity we consider here only the case $E_{\k 1} = E_{\k 2}
= E_{\k}$, as it provides  the main character of the eigenstates.  
Moreover, it is sufficient to consider only the system (\ref{b8}) due to the
double degeneracy of the eigenvalues. 
By standard  method (including the wave function normalization) we obtain the
quasiparticle operators 
\be 
\alpha_{\k} = u_{\k} \frac{1}{\sqrt{2}} \left( f_{\k 1\ua} + f_{\k 1\da}
\right) + v_{\k} \frac{1}{\sqrt{2}} \left( f_{-\k 2\ua}^{\dagger} 
- f_{-\k 2\da} \right) , 
\label{b10} 
\ee 
and 
\be 
\beta_{-\k}^{\dagger} = -v_{\k} \frac{1}{\sqrt{2}} \left( f_{\k 1\ua} + 
f_{\k 1\da} 
\right) + u_{\k} \frac{1}{\sqrt{2}} \left( f_{-\k 2\ua}^{\dagger} 
- f_{-\k 2\da} \right) , 
\label{b11} 
\ee  
where 
\be 
u_{\k} = \frac{1}{\sqrt{2}} \left( 1 + \frac{E_{\k}}{\sqrt{E_{\k}^2+ \Delta^2}}
\right)^{1/2} , ~~~~~
v_{\k} = \frac{1}{\sqrt{2}} \left( 1 - \frac{E_{\k} }{\sqrt{E_{\k}^2+\Delta^2}}
\right)^{1/2} . 
\label{b12} 
\ee   
Again, we have a combination of the two types of states.   
The corresponding wave equation which replaces the Bogolyubov-De Gennes
equation for one-band singlet superconductor reads in the effective-mass
approximation 
\be 
i \hbar \partial_t 
\left( 
\begin{array}{c}
\psi_1 \\
\psi_2 \\
\psi_3 \\
\psi_4
\end{array} 
\right) 
= \left( - \frac{\hbar^2}{2m^{\ast}} \nabla^2 - \mu \right) 
\left( 
\begin{array}{c}
\psi_1 \\
\psi_2 \\
-\psi_3 \\
-\psi_4
\end{array} 
\right)
+ \Delta 
\left( 
\begin{array}{c}
\psi_4 \\
-\psi_3 \\
-\psi_2 \\
\psi_1
\end{array} 
\right)  . 
\label{b13} 
\ee 
We see that the pairing couples explicitly the particle and hole components (
$\psi_1$ with $\psi_4$, $\psi_2$ with $-\psi_3$, etc.). 
This equation  forms a basis for the discussion of inhomogeneous paired
states.   



\begin{references}
\bibitem{15}  The quasiparticle bandwidth $\tilde{W} = Wq$ is much larger than
the coupling constant $Jt^2$, since even close to the Mott-Hubbard localization
$q \simeq 1/2$ and $t^2 \lesssim 0.1$ (cf. Ref.12).
\bibitem{20}  Paranthetically, the {\it asymptotically free} particles
represent state very similar to that of Landau-Fermi liquid, particularly if
the gauge fields are ignored (we make that assumption to simplify the argument
here, as we want to point only the connection to the formalism in the two-band
case in nonrelativistic situation).
\end{references}
\end{document}